\documentclass{stanfest}
\author{Andreas A.\ C.\ Sander}[ARI]
\author{Roel R.\ Lefever}[ARI]
\author{Gemma Gonz\'alez-Tor\`a}[ARI]
\affil[ARI]{Zentrum für Astronomie der Universität Heidelberg, Astronomisches Rechen-Institut, 
   Mönchhofstr. 12-14, 69120 Heidelberg, Germany}

\title{The complex dependencies of Wolf-Rayet winds -- Insights from detailed radiative transfer models}
\headtitle{Insights on Wolf-Rayet winds from detailed radiative transfer models}

\begin{document}

\maketitle

\begin{abstract}
With their emission-line dominated spectra, the appearance of Wolf-Rayet stars is shaped by their strong stellar winds. Yet, the physical mechanisms behind their high mass loss have long remained enigmatic. While we know nowadays that radiative driving is sufficient to explain WR-type outflows,
a coherent description of them is still lacking, not least to the complex physical conditions invalidating some of the approximations sufficient for other hot-star winds.

One promising instrument towards a better understanding of WR winds are comoving-frame, non-LTE stellar atmosphere models including a consistent solution of the hydrodynamics. While so far limited to 1D, their detailed treatment of the radiative transfer and the population numbers is key to overcome the traditional problem of connecting stellar structure models with observed spectra. By creating larger model sequences, we can identify previously unknown scalings and describe trends of WR wind quantities with fundamental stellar parameters and abundances.

This article will present a summary of recent insights on WR-type winds, revealing a complex picture with various remaining challenges. Beside covering classical, hydrogen-free WR stars, we present new results to uncover dependencies of later-type WR stars and the presence of hydrogen-containing envelopes.
We further discuss oncoming challenges and insights from 2D and 3D RHD simulations which need to be mapped into 1D dynamical atmosphere models.
\end{abstract}


\section{Introduction}
\label{sec:intro}

The winds of Wolf-Rayet (WR) stars have been a puzzle since the advent of radiation-driven wind theory. Historically, their large inferred mass-loss rates ($\dot{M}$) gave rise to the so-called ``momentum problem'', meaning that the ``wind efficiency''
\begin{equation}
  \label{eq:eta}
  \eta = \frac{\dot{M}v_\infty}{L/c}
\end{equation}
much larger than unity in WR winds, while it is lower than unity for OB winds. The ratio in Eq.\,(\ref{eq:eta}) balances the wind momentum $\dot{M} v_\infty$ -- $v_\infty$ denoting the terminal wind velocity -- with the luminosity $L$ divided by the speed of light $c$. $\eta$ further denotes the average amount of scattering processes a photon undergoes before leaving the wind. With $\eta > 1$, WR stars are beyond the so-called ``single scattering limit''.
Extreme values of $\eta > 50$ inferred in the 1990s \citep[e.g.,][]{Hamann+1995} are theoretically possible \citep{LucyAbbott1993,Springmann1994}, but essentially all empirical findings so far were later reduced by the inclusion of (optically thin) clumping \citep[e.g.,][]{Hillier1991,HamannKoesterke1998} into the spectral analysis as well as a revision of the luminosities with the advent of Gaia distances. Still, values well above unity remain with typical values of $\eta$ now being around $5$ to $20$ for WRs in the Milky Way \citep{Hamann+2019,Sander+2019}. As such, $\eta = 1$ is not really a limit as e.g.\ pointed out by \citet{FriendCastor1983}. \citet{Gayley+1995} demonstrated that multiple scattering can be highly efficient if sufficient line overlapping exists and concluded that the challenge in understanding WR winds is rather an ``opacity problem'' than a momentum problem. Indeed, Monte Carlo simulations by \citet{SpringmannPuls1998} could demonstrate that while for OB-type winds photons can escape the star at certain wavelengths without any interactions, these ``gaps'' are closed in WR-type winds due to the presence of different ionization stages. For their models representing WN5 stars, these are in particular Fe\,\textsc{iv} to Fe\,\textsc{vi}. Notably, \citet{SpringmannPuls1998} also predicted $\eta < 10$ from their Monte Carlo simulations, which was low at that time, but according to \citet{Hamann+2019} now holds for almost the complete analysed Galactic WN population.

\section{Hydrodynamically-consistent atmosphere models}

The challenge of identifying the necessary opacity has led to a different methodology for the study of WR-type winds that inherently includes multiple scattering.
Beside Monte Carlo calculations, this can also be ensured by performed the radiative transfer in the co-moving frame \citep[CMF,][]{Mihalas+1975,Mihalas+1976}. Modern computers allow to perform these calculations combining continuum and line opacities for a large set of elements and ions, thereby providing an explicit calculation of the flux-weighted opacities $\kappa_F$ and the (here 1D) radiative acceleration
\begin{equation}
  a_\mathrm{rad}(r) = \frac{4\pi}{c} \int\limits^{\infty}_{0} \kappa_\nu H_\nu \mathrm{d}\nu \equiv \frac{\kappa_F L}{4\pi c r^2}\mathrm{.}
\end{equation}

With the such calculated $a_\mathrm{rad}$ being written as a function of radius only, the hydrodynamic equation of motion reads
\begin{equation}
  \label{eq:hydro}
   \left( 1 - \frac{a^2}{v^2} \right) v \frac{\mathrm{d}v}{\mathrm{d}r}	= a_\mathrm{rad}(r) - \frac{G M}{r^2} + 2 a \left(\frac{a}{r} - \frac{\mathrm{d}a}{\mathrm{d}r}\right)
\end{equation}
with $a^2 = a_\mathrm{sound}^2 + v_\mathrm{turb}^2$ being the combined quantity from the (isothermal) sound speed $a_\mathrm{sound}$ and a possible turbulent velocity $v_\mathrm{turb}$. For this Eq.\,(\ref{eq:hydro}), a critical point occurs at $v = a$. If only gas and radiation pressure are taken into account, this is the sonic point. In case of an additional turbulent pressure, i.e., $v_\mathrm{turb} > 0$, this critical point is located beyond the sonic point.

\begin{figure}[!t]
\includegraphics[width=\textwidth]{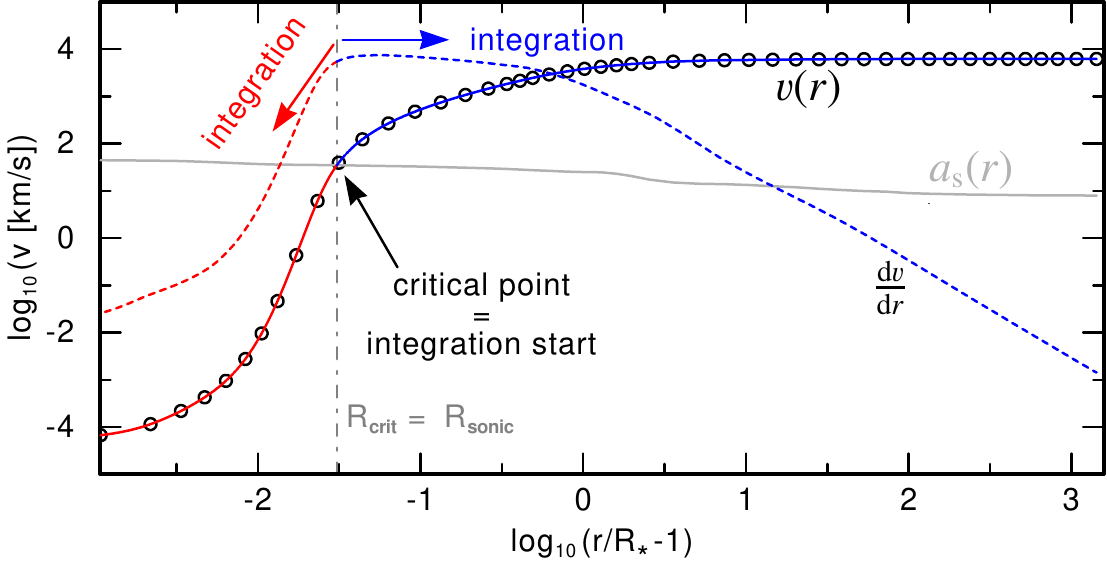}
\caption{A hydrodynamically-consistent solution for the velocity field in a 1D atmosphere model without requiring force multipliers: The grey solid line denotes the sound speed, while the red and blue solid lines show the inward and outward part the resulting velocity field. The corresponding velocity gradients, which are obtained from the equation of motion, are shown as dashed curves of the same color. The black circles mark the course grid of the atmosphere model to which the obtained solution is eventually mapped.}
\label{fig:rkhydro}
\end{figure}

In principle, Eq.\,(\ref{eq:hydro}) can be rewritten such that one obtains an expression for the velocity gradient. Assuming a starting value for $v$ and using the CMF results for $a_\mathrm{rad}$, all quantities are known to obtain $v(r)$ via (numerical) integration. When iterating a stellar atmosphere model with the obtained velocity field, one can eventually obtain a hydrodynamically-consistent solution. The basic idea of this method goes back to \citet{LucySolomon1970}, but the numerical implementation has proven to be tricky \citep[e.g.,][]{Pauldrach+1986}. Moreover, the critical point of Eq.\,(\ref{eq:hydro}) is not to be mixed up with the CAK critical point which is much further out in the wind due to the additional dependence of $a_\mathrm{rad}$ on the velocity gradient. Yet, adapting a CAK-like formulation has found to be very inefficient for dynamically-consistent CMF atmospheres as the necessary calculation of (effective) force multiplier parameters requires doubling the calculation of the radiative transfer, which is usually the most time-consuming part in the iteration process \citep{GraefenerHamann2005,Sander+2017}. Instead, current codes \citep{Sander+2017,Sundqvist+2019} apply the above formulation and obtain the velocity field by inwards and outwards integration of the equation of motion from $v = a$ as illustrated in Fig.\,\ref{fig:rkhydro}. To obtain a sufficient solution for $v(r)$, this iteration has to be performed on a much finer grid than generally used in the model atmosphere and an adaptive step-size Runge-Kutta scheme is employed.
Contrary to the CAK equation of motion (and its extensions), the mass-loss rate $\dot{M}$ does not explicitly appear in Eq.\,(\ref{eq:hydro}). However, it is implicitly fixed from required conservation constraints. As discussed in \citet{Sander+2017}, there are in principle multiple options to express this constraint. The here discussed PoWR$^\textsc{hd}$ code employs a conservation of the total continuum optical depth, which has been a general boundary constraint for PoWR models since \citet{Sander+2015}.

\section{OB- versus WR-type winds}

For studying the driving of stellar winds, the detailed calculation of the individual $\kappa_F$ in the time-consuming CMF radiative transfer is actually a big advantage, as one can precisely identify the depth-dependent contributions of each ion to the total radiative force. This allows to get a deeper insight into different wind regimes and eventually -- by performing calculations with selected parameter changes -- to identify how radiation-driven winds scale when different abundances or stellar parameters change. An inspection of models covering different parameter regimes shows significant changes in the leading elements and ions, e.g.\ when varying the effective temperature of the star. The underlying origin is the different efficiency in line driving connected to the different ionization stages \citep[see, e.g.,][for specific examples]{Sander2023}. Yet, at the wind onset, it is usually iron that provides the necessary line acceleration (on top of the contribution from free electrons) to overcome gravity and launch a stellar wind. Therefore, despite many elements participating in the acceleration of the outer wind, iron -- and thus the initial metallicity of a star -- is crucial for the scaling of $\dot{M}$, even for carbon-rich WC stars \citep{Sander+2020}.

Traditionally, stellar wind studies focussed on considering opacities from the ions and elements visible in the observed spectrum. In the winds of most WR stars, we see essentially the same elements and the same ionization stages as in O stars. In very dense winds, more ionization stages appear in the same spectrum, correctly leading to the assumption that the change in ionization stage is important to explain the nature of WR winds, as discussed above. Yet, this is not enough to understand the onset and the high mass-loss rate of WR-type winds. For this, the layers below the observed surface play a crucial role. The OPAL \citep{IglesiasRogers1991} and Opacity Project \citep{CuntoMendoza1992} efforts prompted the idea that the significant opacity provided by the iron M-shell ions could be responsible for launching the winds of WR-type stars \citep[e.g.,][]{PistinnerEichler1995}. The pioneering calculations by \citet{GraefenerHamann2005} including these higher iron ionization stages in CMF atmosphere models demonstrated that this concept could explain both the mass-loss rate as well as the observed spectrum of a WC star. Yet, the computational resources were restricted. While the developed method helped to get important insights on hydrogen-rich WN stars \citep{GraefenerHamann2008} and the nature of WR winds as a result of the proximity to the Eddington Limit, it could not be scaled to large samples with deep wind launching. These limitations could finally be overcome by the a new generation of dynamically-consistent models \citep{Sander+2020}. Following the approach sketched in Fig.\,\ref{fig:rkhydro}, \citet{SanderVink2020} obtained the first mass-loss recipe for classical, hydrogen-free WR stars based on a large set of dynamically-consistent calculations. In all of these models, the winds are launched from the so-called ``hot iron bump'' arising from the M-shell opacities, implying that -- in a 1D stationary picture -- gravity is already overcome at optical depths usually much larger than unity. In principle, this can also be seen as the main difference between OB- and WR-type winds: In WR-type winds, the critical point (as defined above) is located in the optically thick regime, while this is not the case in OB-type winds. Unfortunately, however, the spectral transition between OB and WR classification does not exactly align with this trend and (moderate) WR-type spectra can occur before $\tau(R_\mathrm{crit})$ surpasses unity.

\begin{figure}[!t]
\includegraphics[width=\textwidth]{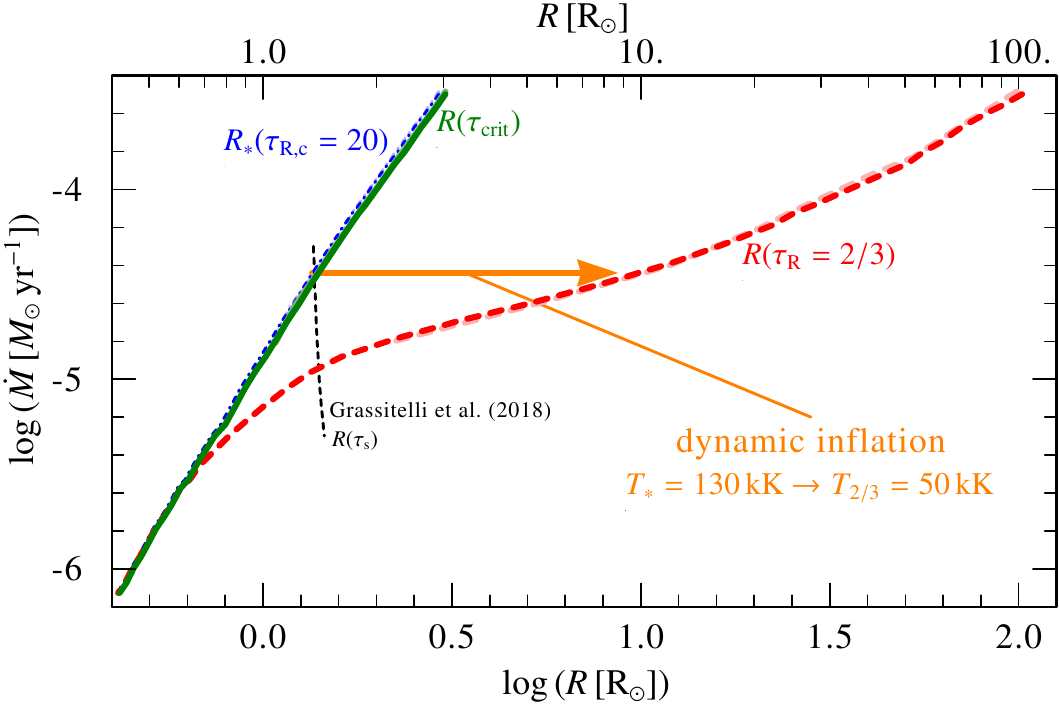}
\caption{Mass-loss rates plotted over different radii for a sequence of hydrodynamically-consistent atmosphere models using a fixed $L$, $M$, and chemical composition: The blue dash-dotted curve denotes the radius at the inner model boundary ($R_\ast$), while the green solid curve shows the radius at the critical point as defined by Eq.\,(\ref{eq:hydro}). The red dashed curve shows the radius of $\tau_\mathrm{Ross} = 2/3$, reflecting where the light in the continuum leaves the star. The black dashed curve shows results from stellar structure calculations with a sonic boundary conditions from \citet{Grassitelli+2018}.}
\label{fig:dyninfl}
\end{figure}
´
\section{Beyond compact Wolf-Rayet stars}

The bulk of WR stars are interpreted as He-burning stars that have been stripped off their outer layers -- either intrinsically or via interaction. In particular for hydrogen-free stars, stellar structure models predict that these stars are rather compact and essentially located at a ``He Main Sequence'' in the Hertzsprung-Russell diagram (HRD). This assumption has also led many investigations, including the efforts from \citet{SanderVink2020} as it reduced the parameter space that needs to be covered. Yet, this reduction can lead to an underestimation of the mass-loss rate as demonstrated more recently by \citet{Sander+2023}. From a series of dynamically-consistent models, they derived that for winds launched by the hot iron bump, the mass-loss rate scales with
\begin{equation}
  \label{eq:mdrcrit}
  \dot{M} \propto R_\mathrm{crit}^3\mathrm{,}
\end{equation}
meaning that the mass-loss rate of (at least compact) WR stars is strongly affected by the precise location of the critical radius. This is also illustrated in Fig.\,\ref{fig:dyninfl}, where radii and mass-loss rates for a sequence of He star models with $20\,M_\odot$ are plotted. In addition, the results from structure calculations with sonic boundary conditions from \citet{Grassitelli+2018} are shown. The crossing of the $R_\mathrm{crit}$-curve with the structure model calculations shows the consistent solution for a hydrogen-free star located exactly on the He zero-age main sequence. The models predict a strong wind with a mass-loss rate of $\sim$$10^{-4.4}\,M_\odot\,\mathrm{yr}^{-1}$ that creates an extended atmosphere effectively cloaking the star with an expanding envelope that will appear as an object with an effective temperature of about $50\,$kK rather than the $130\,$kK representing the corresponding $T_\mathrm{eff}(R_\mathrm{crit})$. At the same time, the apparent radius of the star is about six times larger than the edge of the hydrostatic regime. If $R_\mathrm{crit}$ is even larger, for example due to the contraction of the star when evolving towards the He main sequence, the mass-loss predicted from the models increases even further. For sufficiently large $R_\mathrm{crit}$, even $\dot{M} > 10^{-4}\,M_\odot\,\mathrm{yr}^{-1}$ is theoretically possible as illustrated by a few example model results listed in Table\,\ref{tab:hdlargemdot}. These examples show that such values could even be reached with moderate $L$ and $M$ as long as the wind launching can be shifted sufficiently outwards.

\begin{table}[!t]
\caption{Exemplary results for dynamically-consistent atmosphere hydrogen-free WN-type models resulting in very large mass-loss rates \citep[mainly taken from][]{Sander+2023}.}
\label{tab:hdlargemdot}
\begin{center}
{\small
\begin{tabular}{cccccc}
\hline
	$\log L/L_\odot$ & $M\,[M_\odot]$ & $R_\mathrm{crit}\,[R_\odot]$ & $Z / Z_\odot$  & $\log \left(\dot{M}\,[M_\odot\,\mathrm{yr}^{-1}]\right)$ & $v_\infty [\mathrm{km\,s}^{-1}]$ \\\hline
					     $6.4$         & $60.3$          & $2.7$          &   $1.0$      &   $-3.65$   &  $3400$ \\
				       $5.35$        & $12.9$          & $2.6$          &   $1.0$      &   $-3.68$   &  $605$ \\
				       $5.7$         & $18$            & $2.5$          &   $0.5$      &   $-3.67$   &  $680$ \\
				       $5.7$         & $20$            & $3.6$          &   $1.0$      &   $-3.28$   &  $840$ \\							
\hline
\end{tabular}
}
\end{center}
\end{table}

When comparing to empirically determined mass-loss rates, $\dot{M}$ values as in Table\,\ref{tab:hdlargemdot} appear way too high. Yet, large radii are predicted during very fast stages of stellar evolution, e.g., when a star returns from the red supergiant branch. Hence, the atmosphere calculations illustrate that under the right conditions significant radiatively-driven mass loss can remove one solar mass in a few thousand years, even without considering additional eruptive mass loss. 

\begin{figure}[!t]
\includegraphics[width=\textwidth]{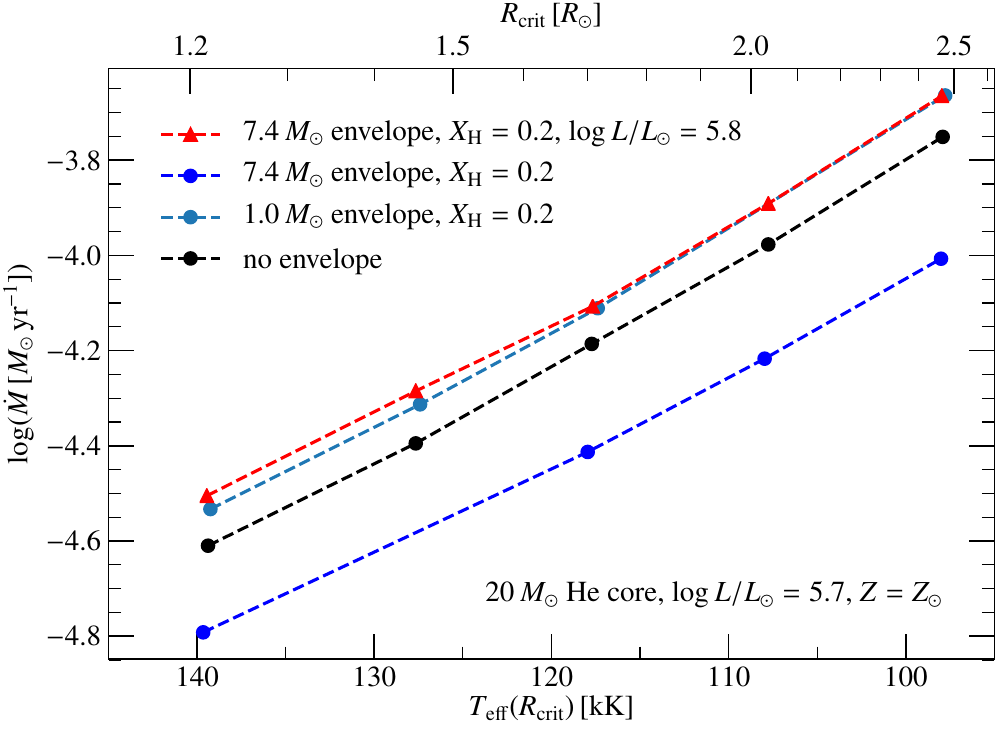}
\caption{Mass-loss rates versus effective temperature at the critical point obtained from sequences of hydrodynamically-consistent atmosphere models assuming different hydrogen-containing envelopes on top of a $20\,M_\odot$ He star. For all models with envelopes, a surface hydrogen mass fraction of $0.2$ (by mass) is assumed. Individual models are shown by dots except for the sequence where an additional shell-burning luminosity of $0.1\,$dex is assumed (red triangles).
The upper x-axis denotes $R_\mathrm{crit}$ in $R_\odot$ for the models with $\log L/L_\odot = 5.7$.}
\label{fig:wrenvseq}
\end{figure}
For longer-lasting stages, the picture is of course much less extreme. In fact, contrary to Eq.\,(\ref{eq:mdrcrit}), the mass-loss rates of some of the later-type WR stars are even lower than those of earlier types \citep[e.g.][]{Hamann+2019}. This illustrates that the WR population as a whole cannot only contain objects with no or a negligible hydrogen envelope like WR\,111 \citep{GraefenerHamann2005} which are intrinsically compact but have very extended winds. In particular, hydrogen-containing WN stars mark a complex population due to their additional envelopes which can significantly vary in both surface abundance and mass. In Fig.\,\ref{fig:wrenvseq}, we show how the mass-loss rate can be affected when adding a hydrogen-containing envelope. Hydrogen provides more free electrons than helium, which slightly increases the resulting mass-loss rates \citep[see also][]{Sander+2020,Sander+2023}. This is the dominating effect for a small envelope mass. When the envelope mass gets larger, the additional gravitational pull overcomes this effect and the resulting change in $\dot{M}$ becomes negative. However, this so far ignores any structural response of the star. The existence of a hydrogen-containing shell increases the radius and eventually can also provide additional luminosity via shell-burning \citep[e.g.][]{Farrell+2020}. These effects are also shown in Fig.\,\ref{fig:wrenvseq}. The increasing trend of the curves with $R_\mathrm{crit}$ highlight that a moderate increase in the (critical) radius can overcome the negative effect of additional mass. Adding $0.1\,$dex to the luminosity totally reverses the picture and yields a generally higher $\dot{M}$.

At first sight, the numerical experiments above seem to be in conflict with empirical mass-loss descriptions for WR stars such as \citet{NugisLamers2000} or \citet{Shenar+2019}, which actually predict an increase of $\dot{M}$ when the surface hydrogen abundance goes down. Yet, this contrast rather implies that for observed samples, the additional mass component must be quite large. In fact, in many cases of late-type WNh stars the effect is propably so large that wind launching from the hot iron bump fails, as suggested by the efforts to model WR\,22 \citep{GraefenerHamann2008}. We can confirm this suspicion with preliminary work on the Quintuplett WN9h star WR 102i (aka LHO99), which has a luminosity of $L = 10^{6.2}\,L_\odot$. From the dynamically-consistent modelling, we require a stellar mass of $\sim$$75\,M_\odot$ (Lefever et al., in prep.) to obtain the necessary $\dot{M} = 10^{-4.76}\,M_\odot\,\mathrm{yr}^{-1}$ to fit the spectral appearance in the $K$ band. For the given luminosity, a He-burning star would only have $\sim$$43\,M_\odot$ \citep[using the relation from][]{Graefener+2011}, meaning that the star has a huge hydrogen envelope. Yet, the star is likely no longer in the stage of central hydrogen burning as the determined hydrogen surface mass fraction is only $\sim$$0.45$. Moreover, a chemically homogeneous structure for the inferred luminosity would also imply a higher mass of $\sim$$86\,M_\odot$.

Turning all of these effects into a new mass-loss recipe is ongoing work as the results are not trivial with respect to scaling. For example, the models with additional shell-burning shown in Fig.\,\ref{fig:wrenvseq} have a $0.04$\,dex lower $L/M$-ratio than the envelope-free models, but they still yield a higher $\dot{M}$. For the hydrogen-free sample, the formulae of \citet{SanderVink2020} and \citet{Sander+2023} can be combined into a description that depends on $L/M$, $Z$, and $R_\mathrm{crit}$ or $T_\mathrm{eff}(R_\mathrm{crit})$. A larger parameter study will be required to extract whether this description can be extended further, but an additional $M$- and $X_\mathrm{H}$-dependency \citep[see also][]{Sander2024} is likely unavoidable. Moreover, the existence of a ``failed wind regime'' in atmosphere and structure models \citep{Grassitelli+2018,Sander+2023} could mean that no continous $\dot{M}$-description exists covering all kinds of WR stars, but this conclusion might be preliminary as the solution topology could significantly be affected when considering turbulent pressure as we will motivate below. 


\section{Open issues and the need for including multi-dimensional effects}

With their local consistency and the detailed calculation of opacities, the new generation of hydrodynamically-consistent atmosphere models has set new standards in the prediction of mass-loss rates for massive stars. In particular for WR stars, the studies have revealed a previously unexpected drop in the mass-loss rates when the winds start to become optically thin, which helps tremendously to understand the sudden disappearance of massive WR-type objects below a certain, metallicity-dependent, luminosity cut-off as well as the disappearance of WR signatures at lower metallicity \citep[see also][]{SanderVink2020,Shenar+2020}. 

Yet, several challenges remain when trying to explain the wide range of observed WR stars. One of these issues is the high amount of clumping required in WR models to reach the observed terminal velocities. Interestingly, this effect is opposite to what is obtained in similar models for OB-type stars \citep{Bjoerklund+2021}, hinting that there might be missing physics in the current 1D model approaches which could affect the derived terminal velocities and possibly also the location of the critical point. Nowadays, multi-dimensional simulations \citep[e.g.,][see also these proceedings]{Moens+2022} start to be possible, but they still have to make considerable approximations. For the foreseeable future, a combination of 3D and 1D calculation efforts is the most promising strategy where new insights obtained in 2D and 3D simulations are meaningfully parametrized in 1D models to explore a large parameter space. One prominent example that needs to be investigated is the occurrence of radiatively driven turbulence in the case of ``failed'' wind launching from the hot iron bump. This phenomenon has recently been seen in new 2D simulations of O (super-)giants by \citet[see also these proceedings]{Debnath+2024} and is likely to occur also in some WR stars, e.g., where the observed spectra cannot be reproduced with deep-launching models, either because they fail to obtain a solution \citep[see][]{Sander+2023} or because they predict too strong emission lines. Further aspects where new insights emerge from multi-D modelling, partially also underpinned with observational evidence, concern the treatment of wind clumping itself, the existence of a density-velocity anti-correlation, or the existence of material with a much higher ionization stage than observed in the main spectral diagnostics. Together with the so far usually ignored structural constraints discussed above, a new momentum for re-considering some of the fundamental treatments of stellar atmosphere modelling starts to emerge with a significant potential to not only alter our picture of WR winds, but reshape our perception of the properties of massive stars in general.


%
\acknowledgements{AACS and RRL are supported by the Deutsche Forschungsgemeinschaft
(DFG -- German Research Foundation) in the form of an Emmy Noether Research
Group -- Project-ID 445674056 (SA4064/1-1, PI Sander). 
GGT acknowledges funding from the Deutsche Forschungsgemeinschaft (DFG -- German Research Foundation)
Project-ID 496854903 (SA4064/2-1, PI Sander).
This work made extensive use of NASA's Astrophysics Data System (ADS).}

%
\bibliographystyle{stanfest_bibstyle}
\bibliography{sanderbib}

\end{document}